\documentclass[prd,
twocolumn,
nofootinbib,
showpacs ,amssymb, aps]{revtex4}
\usepackage{graphicx}
\input{epsf}

\usepackage{amsmath,amssymb}
\usepackage{bm}

\newcommand{\dalm}{\kern1pt\vbox{\hrule height 0.9pt\hbox{\vrule width 0.9pt
\hskip 2.5pt\vbox{\vskip 5.5pt}\hskip 3pt\vrule width 0.3pt}\hrule height 0.3pt}
\kern1pt}

\usepackage{natbib}
\usepackage{wasysym}

\newcommand{\gt}{\tilde{g}}

\newcommand{\nablat}{\tilde{\nabla}}

\newcommand{\Aether}{\AE ther }

\begin{document}

\title{Black holes and neutron stars in the generalized tensor-vector-scalar theory}

\author{Paul D. Lasky}
	\email{lasky@tat.physik.uni-tuebingen.de}
	\affiliation{Theoretical Astrophysics, Eberhard Karls University of T\"ubingen, T\"ubingen 72076, Germany}

		\begin{abstract}
			Bekenstein's Tensor-Vector-Scalar (TeVeS) theory has had considerable success as a relativistic theory of Modified Newtonian Dynamics (MoND).  However, recent work suggests that the dynamics of the theory are fundamentally flawed and numerous authors have subsequently begun to consider a generalization of TeVeS where the vector field is given by an Einstein-\Aether action.  Herein, I develop strong-field solutions of the generalized TeVeS theory, in particular exploring neutron stars as well as neutral and charged black holes.  I find that the solutions are identical to the neutron star and black hole solutions of the original TeVeS theory, given a mapping between the parameters of the two theories, and hence provide constraints on these values of the coupling constants.  I discuss the consequences of these results in detail including the stability of such spacetimes as well as generalizations to more complicated geometries.
		\end{abstract}
		
		\received{}\published{}
		

		\maketitle

\section{Introduction}
Modified Newtonian Dynamics (MoND) \cite{milgrom83} has enjoyed significant success as a phenomenological model of gravity without the requirement for dark matter (for a review see \cite{sanders02}).  However, the discovery of a complete relativistic theory generalizing MoND has proven extremely difficult.  Recently, Bekenstein's Tensor-Vector-Scalar (TeVeS) theory \cite{bekenstein04} has received considerable attention; in the weak acceleration limit TeVeS  has been shown to reproduce MoND, and in the Newtonian limit the parametrized post-Newtonian coefficients, $\beta$ and $\gamma$, are consistent within limits imposed by solar system experiments (for a comprehensive review of TeVeS see \cite{skordis09}).  

Despite its success, there exists mounting evidence that TeVeS suffers from dynamical problems.  Firstly, \citet{seifert07} showed that the Schwarzschild-TeVeS solution \cite{giannios05} is unstable to linear perturbations for experimentally and phenomenologically valid values of the various coupling parameters.  By analysing a broad variety of dynamical situations, \citet{contaldi08} then showed that the vector field is prone to the formation of caustics analogously to Einstein-\Aether theories where the vector field is given by a Maxwellian action \cite{jacobson01}.  Finally, \citet{sagi09} has shown that the vector field is constrained by the cosmological value of the scalar field in such a way that it prevents the scalar field from evolving and also forces it to be negative, thereby inducing superluminal propagation of scalar waves.  \citet{contaldi08} and \citet{skordis08} have provided a generalization of TeVeS, whereby the vector field action is not given by the Maxwellian form of TeVeS, but a more general Einstein-\Aether form.  In this way the vector field action is the ``\ldots most general diffeomorphism invariant action, which is quadratic in derivatives and consistent with the $A^{2}=-1$ constraint'' \cite{contaldi08}, whereas the tensor and scalar field actions are identical to those of TeVeS.  This type of generalization to the vector field action is motivated from Einstein-\Aether theories, where it was shown to stabilize dynamical problems.  Indeed, this generalization of TeVeS has been shown to resolve the caustic formation problem present in the original TeVeS theory \cite{contaldi08}.

Calculations of various quantities in generalized TeVeS are still in their infancy.  \citet{skordis08} presented the cosmological equations for the theory and showed that they are identical with the TeVeS equations up to a rescaling of Hubble's constant.  \citet{sagi09} has looked at the PPN parameters within this theory and found them to not conflict with solar system experiments for suitable values of the various parameters of the theory.  I continue investigations by exploring strong-field solutions, finding the spherically symmetric, static solutions of the field equations for neutron stars, neutral and charged black holes (sections \ref{NS} and \ref{charge}).  Interestingly, I find all these solutions to be the same as for the original TeVeS theory given a simple substitution of the vector field coupling parameters.  This implies that the work of \citet{giannios05} for neutral black holes in TeVeS, \citet{sagi08} for charged black holes and \citet{lasky08} for neutron stars is also correct for the generalized TeVeS theory, given the aforementioned substitution of parameters.  Spherically symmetric neutron stars and black holes are therefore observationally indistinguishable in TeVeS and generalized TeVeS.  In section \ref{conc}, I discuss in detail further implications of this work, including generalizations to more complex spacetimes and stability arguments.

\section{Generalized TeVeS field equations}\label{GenEquations}
Generalized TeVeS \cite{skordis08,contaldi08}, as with the original TeVeS theory \cite{bekenstein04}, is built upon three gravitational fields; an Einstein tensor field, $g_{\mu\nu}$, a normalized timelike vector field, $A^{\mu}$, and a dynamical scalar field, $\varphi$.  These three objects combine in the following way to give the physical metric;
\begin{align}
		\gt_{\mu\nu}=&{\rm e}^{-2\varphi}\left(g_{\mu\nu}+A_{\mu}A_{\nu}\right)-{\rm e}^{2\varphi}A_{\mu}A_{\nu},
\end{align}
which is the metric measured by clocks and rods.  Throughout this article, quantities with and without tildes are measured in the physical and Einstein frames respectively.  
%For example, the vector field is raised and lowered using the Einstein tensor, $A_{\mu}\equiv g_{\mu\alpha}A^{\alpha}$.  
%The original TeVeS theory has two coupling constants which represent the scalar field and vector field couplings, $k$ and ${\cal K}$ respectively.  Generalizing TeVeS  from the Maxwellian vector field to an Einstein-\Aether vector field implies one now has four vector field coupling constants, $K$, $K_{+}$, $K_{2}$ and $K_{4}$, while also retaining the scalar field coupling
The original and generalized versions of the theory differ in the vector field action.  Bekenstein's version had a Maxwellian action, and subsequently one vector field coupling constant, ${\cal K}$.  The generalized version has an Einstein-\Aether vector field action with four associated coupling constants, $K$, $K_{+}$, $K_{2}$ and $K_{4}$\footnote{I follow the notation of \citet{sagi09}, which is different to that of \citet{skordis08} who denotes the four parameters by $K_{B}$, $K_{+}$, $K_{0}$ and $K_{A}$.  These are related as follows; $K\equiv K_{B}\equiv(c_{1}-c_{3})/2$, $K_{+}\equiv K_{+}\equiv(c_{1}+c_{3})/2$, $K_{2}\equiv K_{0}\equiv c_{2}$ and $K_{4}\equiv K_{A}\equiv -c_{4}$, where the $c_{i}$ are those of the Einstein-\AE ther theory (see \citet{jacobson08}).}.  Bekenstein's original TeVeS theory is recovered by letting $K_{+}=K_{2}=K_{4}=0$ and $K={\cal K}$.  

The modified Einstein field equations can be expressed analogously to TeVeS as
\begin{align}
	G_{\mu\nu}=8\pi G\left[\tilde{T}_{\mu\nu}+\left(1-{\rm e}^{-4\varphi}\right)A^{\alpha}\tilde{T}_{\alpha(\mu}A_{\nu)}+\tau_{\mu\nu}\right]+\Theta_{\mu\nu}.\label{Einstein}
\end{align}
Here, $G_{\mu\nu}$ is the Einstein tensor associated with the Eintein frame and $\tilde{T}_{\mu\nu}$ is the physical stress-energy tensor.  The tensor, $\tau_{\mu\nu}$, has the same definition as that in the original theory
\begin{align}
	\tau_{\mu\nu}&:=\frac{\mu}{kG}\Big[\nabla_{\mu}\varphi\nabla_{\nu}\varphi-\frac{1}{2}g^{\alpha\beta}\nabla_{\alpha}\varphi\nabla_{\beta}\varphi g_{\mu\nu}-A^{\alpha}\nabla_{\alpha}\varphi\nonumber\\
	&\times\Big(A_{(\mu}\nabla_{\nu)}\varphi-\frac{1}{2}A^{\beta}\nabla_{\beta}\varphi g_{\mu\nu}\Big)\Big]-\frac{\mathcal{F\left(\mu\right)}}{2k^{2}\ell^{2}G}g_{\mu\nu},
\end{align}
where $k$ is the scalar field coupling constant, $\ell$ is a positive scalar of dimension length, $\mu$ is a function associated with the MoND acceleration scale, and $\mathcal{F}\left(\mu\right)$ is a function that is not predicted by the theory.  Only the strong-field limit of the theory is considered in this article, whereby an excellent approximation is $\mu=1$ (for more details see %the discussions in
\cite{bekenstein04,giannios05,sagi08,contaldi08}).  Moreover, in this case one finds that 
%the free function 
$\mathcal{F}$ diverges logarithimcally, but is exactly cancelled out in the field equations implying one is free to ignore this function %when working in the strong-field limit 
\cite{contaldi08}.  It is worth noting that $\mathcal{F}$ only diverges in the strong-field limit considered herein, implying this divergence is not a general concern for the theory.  

As mentioned, the extra degrees of freedom attributed to the generalized version of TeVeS appear in the vector-field action, which manifests itself partly in the $\Theta_{\mu\nu}$ term of equation (\ref{Einstein}) (for details of these derivations see \cite{skordis08,contaldi08}).  For clarity in the expressions, 
$\Theta_{\mu\nu}$ can be expressed as a linear sum of terms associated with each coupling parameter, plus one term associated with the Lagrange multiplier, $\lambda$;
\begin{align}
	\Theta_{\mu\nu}:=\Theta^{K}_{\mu\nu}+\Theta^{K_{+}}_{\mu\nu}+\Theta^{K_{2}}_{\mu\nu}+\Theta^{K_{4}}_{\mu\nu}+\Theta^{\lambda}_{\mu\nu},
\end{align}
where
\begin{align}
	\Theta^{K}_{\mu\nu}:=&K\left(F_{\alpha\mu}{F^{\alpha}}_{\nu}-\frac{1}{4}g_{\mu\nu}F_{\alpha\beta}F^{\alpha\beta}\right),
\end{align}
\vspace{-0.6cm}
\begin{align}
	\Theta^{K_{+}}_{\mu\nu}:=K_{+}&\Big[S_{\mu\alpha}{S_{\nu}}^{\alpha}-\frac{1}{4}g_{\mu\nu}S_{\alpha\beta}S^{\alpha\beta}\nonumber\\
		&+\nabla_{\alpha}\left(A^{\alpha}S_{\mu\nu}-{S^{\alpha}}_{(\mu}A_{\nu)}\right)\Big],
\end{align}
\vspace{-0.6cm}
\begin{align}
	\Theta^{K_{2}}_{\mu\nu}:=K_{2}&\Big[g_{\mu\nu}\nabla_{\alpha}\left(A^{\alpha}\nabla_{\beta}A^{\beta}\right)-A_{(\mu}\nabla_{\nu)}\left(\nabla_{\alpha}A^{\alpha}\right)\nonumber\\
		&-\frac{1}{2}g_{\mu\nu}\nabla_{\alpha}A^{\alpha}\nabla_{\beta}A^{\beta}\Big],
\end{align}
\vspace{-0.6cm}
\begin{align}
	\Theta^{K_{4}}_{\mu\nu}:=K_{4}&\Big[\dot{A}_{\mu}\dot{A}_{\nu}+\dot{A}_{\alpha}A_{(\mu}\nabla_{\nu)}A^{\alpha}\nonumber\\
		&-\nabla_{\alpha}\left(\dot{A}^{\alpha}A_{\mu}A_{\nu}\right)-\frac{1}{2}g_{\mu\nu}\dot{A}_{\alpha}\dot{A}^{\alpha}\Big],
\end{align}
\vspace{-0.6cm}
\begin{align}	
	\Theta^{\lambda}_{\mu\nu}:=&-\lambda A_{\mu}A_{\nu}.\label{ThetaLambda}
\end{align}
Here, $F_{\mu\nu}:=\nabla_{[\nu}A_{\mu]}$, $S:=\nabla_{(\nu}A_{\mu)}$
%Here, $F_{\mu\nu}:=\nabla_{\nu}A_{\mu}-\nabla_{\mu}A_{\nu}$, $S_{\mu\nu}:=\nabla_{\nu}A_{\mu}+\nabla_{\mu}A_{\nu}$ 
and $\dot{A}_{\mu}:=A^{\alpha}\nabla_{\alpha}A_{\mu}$.  
%Moreover, $\lambda$ is the Lagrange multiplier which is used to ensure the vector field is timelike and normalized.  
%
Additional to the modified Einstein equations, one has the vector field equation given by  
\begin{align}
	K&\nabla_{\alpha}F^{\mu\alpha}+K_{+}\nabla_{\alpha}S^{\alpha\mu}+K_{2}\nabla^{\mu}\left(\nabla_{\alpha}A^{\alpha}\right)+\lambda A^{\mu}\nonumber\\
		&+K_{4}\left[\nabla_{\alpha}\left(\dot{A}^{\mu}A^{\alpha}\right)-\dot{A}^{\alpha}\nabla^{\mu}A_{\alpha}\right]+\frac{8\pi\mu}{k} A^{\alpha}\nabla_{\alpha}\varphi g^{\mu\beta}\nabla_{\beta}\varphi\nonumber\\
	&=8\pi G\left(1-{\rm e}^{-4\varphi}\right)g^{\mu\alpha}\tilde{T}_{\alpha\beta}A^{\beta},\label{vector}
\end{align}
which again reduces to the original TeVeS vector field equation when $K_{+}=K_{2}=K_{4}=0$ and $K={\cal K}$.  Finally, the scalar field equation is
\begin{align}
	\nabla_{\beta}&\left[\mu\left(g^{\alpha\beta}-A^{\alpha}A^{\beta}\right)\nabla_{\alpha}\varphi\right]\nonumber\\
		&\qquad=kG\left[g^{\alpha\beta}+\left(1+{\rm e}^{-4\varphi}\right)A^{\alpha}A^{\beta}\right]\tilde{T}_{\alpha\beta},\label{scalar}
\end{align}
which is equivalent to that of the original TeVeS theory.

\section{Neutron Stars and Neutral Black Holes}\label{NS}
\citet{giannios05} first considered static, spherically symmetric vacuum solutions of the original TeVeS field equations.  He found two branches of solutions; one where the vector field has only a non-zero timelike component, and one where the vector field has both non-zero radial and temporal components.  \citet{lasky08} then considered spherically symmetric neutron star solutions of the original TeVeS field equations where the vector field only has a non-zero temporal component.  The black hole solution of \citet{giannios05} can be recovered from the neutron star solutions of \citet{lasky08} by simply letting the matter variables vanish throughout the spacetime.  I therefore begin here by deriving the neutron star solutions and subsequently the black hole solutions.  Moreover, throughout this section I consider a vector field which points in the temporal direction, for which \citet{bekenstein04} has shown support for the general case of static spacetimes (for a discussion of generalizing this work to a more general vector field {\it ansatz} see section \ref{conc}).

A spherically symmetric, static spacetime can be expressed in the Einstein frame as
\begin{align}
	g_{\alpha\beta}dx^{\alpha}dx^{\beta}=-{\rm e}^{\nu(r)}dt^{2}+{\rm e}^{\zeta(r)}dr^{2}+r^{2}d\Omega^{2},\label{EinsteinMetric}
\end{align}
%where $ds^{2}:=g_{\alpha\beta}dx^{\alpha}dx^{\beta}$
%, $\nu=\nu(r)$, $\zeta=\zeta(r)$ 
where $d\Omega^{2}:=d\theta^{2}+\sin^{2}\theta d\phi$.  As mentioned, the vector field is chosen such that is has only a non-zero temporal component.  Normalization of this implies
\begin{align}
	A^{\mu}={\delta^{\mu}}_{t}{\rm e}^{-\nu/2},\label{VectorAnsatz}
\end{align}
further implying that the physical line element is 
\begin{align}
	\tilde{g}_{\alpha\beta}dx^{\alpha}dx^{\beta}=-{\rm e}^{\nu+2\varphi}dt^{2}+{\rm e}^{-2\varphi}\left({\rm e}^{\zeta}dr^{2}+r^{2}d\Omega^{2}\right),
\end{align}
where symmetries also imply $\varphi=\varphi(r)$.  
%Note that this line element is no longer asymptotically flat, providing the cosmological value of the scalar field is non-zero.

Stress-energy tensors in TeVeS and its generalization are defined in terms of the physical frame.  Therefore, a perfect fluid takes the form
\begin{align}
	\tilde{T}_{\mu\nu}=\left(\tilde{\rho}+\tilde{P}\right)\tilde{u}_{\mu}\tilde{u}_{\nu}+\tilde{P}\tilde{g}_{\mu\nu},
\end{align}
where $\tilde{u}_{\mu}$, $\tilde{\rho}$ and $\tilde{P}$ are respectively the four-velocity, energy-density and pressure of the fluid as measured in the physical frame.  
%As the spacetime is static, the four-velocity of the fluid only points in the temporal direction, and hence is aligned with the timelike vector field $A^{\mu}$.  In this case, \citet{bekenstein04} has shown $\tilde{u}_{\mu}={\rm e}^{\varphi}A_{\mu}$, and furthermore the stress-energy tensor can be expressed as
%\begin{align}
%	\tilde{T}_{\mu\nu}={\rm e}^{2\varphi}\tilde{\rho}A_{\mu}A_{\nu}+{\rm e}^{-2\varphi}\tilde{P}\left(g_{\mu\nu}+A_{\mu\nu}\right).
%\end{align}
Conservation of the stress-energy tensor is given by $\tilde{\nabla}_{\alpha}\tilde{T}^{\alpha}{}_{\mu}=0$,
where $\tilde{\nabla}_{\mu}$ is the unique metric connection associated with the physical metric. 
%$\tilde{g}_{\mu\nu}$
The radial component of this equation implies the pressure gradient is expressed as
\begin{align}
	-\tilde{P}'=\left(\frac{\nu'}{2}+\varphi'\right)\left(\tilde{\rho}+\tilde{P}\right),\label{Euler}
\end{align}
where a prime denotes differentiation with respect to the radial coordinate.

The scalar field equation (\ref{scalar}) can be expressed in terms of the metric coefficients, and once integrated to give 
%\cite{lasky08}
\begin{align}
	\varphi'=\frac{kGM_{\varphi}}{4\pi r^{2}}{\rm e}^{\left(\zeta-\nu\right)/2},
\end{align}
where the scalar mass function, $M_{\varphi}(r)$, has been defined as \cite{bekenstein04}
\begin{align}
	M_{\varphi}(r):=4\pi\int_{\hat{r}=0}^{r}\left(\tilde{\rho}+3\tilde{P}\right){\rm e}^{\nu/2+\zeta/2-2\varphi}\hat{r}^{2}d\hat{r}.
\end{align}
%Note that to find a black hole solution of these equations, $\tilde{\rho}=\tilde{P}=0$ implying $M_{\varphi}$ becomes a constant.  
Given the symmetries of the problem and the vector field {\it ansatz}, the radial component of the vector field equation is now trivially satisfied, and the temporal component provides an equation for the Lagrange multiplier, $\lambda$.  Therefore, this is essentially a periphery equation as the Lagrange multiplier is used in the modified field equations in the $\Theta_{\mu\nu}^{\lambda}$ term of equation (\ref{ThetaLambda}).  
%This term is expressed in the Appendix, equation (\ref{applambda}).  
The only remaining equations are the modified Einstein field equations (\ref{Einstein}).  
%The individual components of the Einstein equations are also given in Appendix \ref{comp}.  
After much work, the $tt$, $rr$ and $\theta\theta$ components can be respectively shown to be 
\begin{align}
	r\zeta'+{\rm e}^{\zeta}-1&=8\pi G\tilde{\rho}{\rm e}^{\zeta-2\varphi}r^{2}+\frac{r^{2}}{4}\left(K+K_{+}-K_{4}\right)\nonumber\\
		&\times\left(\frac{\nu'^{2}}{2}-\nu'\zeta'+2\nu''+\frac{4\nu'}{r}\right)+\frac{4\pi r^{2}}{k}\varphi'^{2},\label{EFEcalca}
\end{align}
\begin{align}
	r\nu'+{\rm e}^{\zeta}+1=&8\pi G{\rm e}^{\zeta-2\varphi}\tilde{P}r^{2}-\left(K+K_{+}-K_{4}\right)\frac{r^{2}}{8}\nu'^{2}\nonumber\\
		&+\frac{4\pi r^{2}}{k}\varphi'^{2},
\end{align}
%\vspace{-0.5cm}
\begin{align}
	\frac{r}{4}&\left(2\nu'-2\zeta'-r\nu'\zeta'+2r\nu''+r\nu'^{2}\right)=8\pi G{\rm e}^{\zeta-2\varphi}\tilde{P}r^{2}\nonumber\\
		&\qquad+\frac{r^{2}}{8}\left(K+K_{+}-K_{4}\right)\nu'^{2}-\frac{4\pi r^{2}}{k}\varphi'^{2}.\label{EFEcalcc}
\end{align}
A striking similarity between this system of equations and the equivalent system in the original TeVeS theory presented in \citet{lasky08}, their equations (21-23), is now observed.  In fact, the system of equations presented for the original TeVeS theory is exactly the same, where
\begin{align}
	K+K_{+}-K_{4}={\cal K},\label{map}
\end{align}
in the above equations.  That is, given the neutron star solution of the original TeVeS theory, one derives the solution of generalized TeVeS by making the substitution (\ref{map}).  The spacetime of neutron stars in both theories is therefore identical.  Moreover, allowing the matter terms to vanish in equations (\ref{Euler}-\ref{EFEcalcc}), and therefore recovering the equations governing the generalized TeVeS Schwarzschild spacetime, one can show that the structure of the spacetime is exactly equivalent to that presented in \citet{giannios05} for the original TeVeS theory.  To show this, one must first perform a coordinate transformation given by $r=r(R)$ such that $r={\rm e}^{\xi(R)/2}R$ to put the system in the same set of coordinates used by Giannios.  Then, allowing the matter terms to vanish, the system of equations becomes exactly equivalent to equations (28-30) of \citet{giannios05}, given the expression in (\ref{map}).  That is, the structure of black hole spacetimes, under the symmetry assumptions posed hitherto, is exactly the same as that of the orignial TeVeS theory.  

There is no need to further derive solutions of these systems of equations for both neutron stars and black holes, as the above result implies the results of \citet{lasky08} and \citet{giannios05} hold for the generalized TeVeS theory, providing equation (\ref{map}) is utilized.  One observes from equations (\ref{Euler}-\ref{EFEcalcc}) that the individual coupling constants do not appear independently of one another, but rather neutron stars and black holes can only be used to constrain the combination of parameters, $K+K_{+}-K_{4}$.  As with \citet{lasky08}, we can use observations of neutron star masses to constrain these coupling constants.  In that paper, we gave a conservative estimate of ${\cal K}\lesssim1$ based on the existence of neutron stars with masses of at least $\sim1.5 M_{\odot}$.  Therefore, using the same conservative estimate the combination of the vector coupling constants is constrained to $K+K_{+}-K_{4}\lesssim1$.

In some ways the relationship between the two TeVeS theories given in (\ref{map}) is to be expected.  As was discussed in section \ref{GenEquations}, one recovers the original TeVeS theory by setting $K_{+}=K_{2}=K_{4}=0$ and $K={\cal K}$, which is an obvious subset of solutions offered by the above equation.  However, (\ref{map}) does not imply that this has to be the case, but rather any combination of $K$, $K_{+}$ and $K_{4}$ may hold such that (\ref{map}) is true.  Moreover, these spacetimes are completely independent of the parameter $K_{2}$.  I discuss the consequences of these result in considerably more detail in section \ref{conc}.  

\section{Charged Black Holes}\label{charge}
\citet{sagi08} considered spherically symmetric, charged black holes in the original TeVeS theory, where the vector field is again aligned with the temporal direction.  In this section I reproduce those calculations in the context of the generalized TeVeS theory.  

The {\it ansatz} and the symmetries applied herein imply the metric and vector field are given by equations (\ref{EinsteinMetric}) and (\ref{VectorAnsatz}) respectively.  The stress-energy tensor is now taken to be that of an Einstein-Maxwell field as measured in the physical frame;
\begin{align}
	\tilde{T}_{\mu\nu}=\frac{1}{4\pi}\left(\tilde{\cal F}_{\alpha\mu}\tilde{\cal F}^{\alpha}{}_{\nu}-\frac{1}{4}\tilde{g}_{\mu\nu}\tilde{\cal F}_{\alpha\beta}\tilde{\cal F}^{\alpha\beta}\right),\label{EMstress}
\end{align}
where $\tilde{\cal F}_{\mu\nu}$ is the electromagnetic Faraday tensor in the physical frame. 
Conservation of stress-energy then leads to the Maxwell equations, $\tilde{\nabla}_{\alpha}\tilde{\cal F}^{\mu\alpha}=0$.
The symmetries of the problem imply the only non-zero contribution to the Faraday tensor 
is the $\tilde{\cal F}_{rt}=-\tilde{\cal F}_{tr}$ component \cite{sagi08}.  Moreover, only the temporal component of Maxwell's equations is non-zero, which is once integrated to give
\begin{align}
	\tilde{\cal F}_{rt}=\frac{\tilde{\mathcal{Q}}}{r^{2}}{\rm e}^{2\varphi+\left(\nu+\zeta\right)/2}.
\end{align}
Here, $\tilde{\mathcal{Q}}$ is a constant of integration which is the charge of the black hole.  As with section \ref{NS}, 
%the vector field has been normalized implying 
the only non-zero contribution from the vector field equation (\ref{vector}) is from the temporal component, which gives the Lagrange multiplier expressed as a function of the metric coefficient.  
%This equation is given in the Appendix as equation (\ref{applambdacharge}).  
Evaluating the modified Einstein equations (\ref{Einstein}) and substituting the Lagrange multiplier back through gives the $tt$, $rr$ and $\theta\theta$ components respectively as
\begin{align}
	r\zeta'+{\rm e}^{\zeta}-1=&\frac{G\tilde{\mathcal{Q}}^{2}}{r^{2}}{\rm e}^{\zeta+2\varphi}%\nonumber\\
		+\frac{r^{2}}{4}\left(K+K_{+}-K_{4}\right)\nonumber\\
		&\times\left(\frac{\nu'^{2}}{2}-\nu'\zeta'+2\nu''+\frac{4\nu'}{r}\right)+\frac{4\pi r^{2}}{k}\varphi'^{2},\label{EFEcalcaEM}
\end{align}
\vspace{-0.5cm}
\begin{align}
	r\nu'+{\rm e}^{\zeta}+1=&-\frac{G\tilde{\mathcal{Q}}^{2}}{r^{2}}{\rm e}^{\zeta+2\varphi}-\left(K+K_{+}-K_{4}\right)\frac{r^{2}}{8}\nu'^{2}\nonumber\\
		&+\frac{4\pi r^{2}}{k}\varphi'^{2},
\end{align}
\vspace{-0.5cm}
\begin{align}
	\frac{r}{4}&\left(2\nu'-2\zeta'-r\nu'\zeta'+2r\nu''+r\nu'^{2}\right)=\frac{G\tilde{\mathcal{Q}}^{2}}{r^{2}}{\rm e}^{\zeta+2\varphi}\nonumber\\
		&\qquad+\frac{r^{2}}{8}\left(K+K_{+}-K_{4}\right)\nu'^{2}-\frac{4\pi r^{2}}{k}\varphi'^{2}.\label{EFEcalccEM}
\end{align}
Moreover, the scalar field equation (\ref{scalar}) can be shown to reduce to
\begin{align}
	\left[\varphi'r^{2}{\rm e}^{\left(\nu-\zeta\right)/2}\right]'=\frac{kG\tilde{\mathcal{Q}}^{2}}{4\pi r^{2}}{\rm e}^{2\varphi+\left(\nu+\zeta\right)/2}.\label{scalarEM}
\end{align}
\citet{sagi08} derived the same set of equations in the original TeVeS theory, although they used isotropic coordinates as opposed to Schwarzschild coordinates.  Performing the same coordinate transformation as section \ref{NS}, i.e. $r=r(R)$ such that $r={\rm e}^{\xi(R)/2}R$ where the Einstein metric has component $g_{RR}={\rm e}^{\xi(R)}$, and also using equation (\ref{map}), one finds that equations (\ref{EFEcalcaEM}-\ref{scalarEM}) given above are identical to equations (49-52) of \citet{sagi08}.  That is, given the charged black hole solutions of \citet{sagi08}, one can derive the solution of generalized TeVeS simply making the substitution (\ref{map}), implying the spacetimes in the two theories are identical.

\section{Implications and Generalizations}\label{conc}
Herein I have shown that solutions of the field equations within Bekenstein's original TeVeS framework for spherically symmetric neutron stars \cite{lasky08}, neutral black holes \cite{giannios05} and charged black holes \cite{sagi08} are exactly the same solutions for the generalized TeVeS theory \cite{skordis08,contaldi08}, given the substitution (\ref{map}).  An immediate question one must ask is; {\it how much can these results be generalized?}  That is, if more complex solutions of the original TeVeS field equations are found, will the same solution of the generalized TeVeS equations exist given the substitution (\ref{map})?  The answer to this question is a simple ``no''.  In fact, a quick calculation reveals that this is not even the case in more complex spherically symmetric spacetimes.  For example, consider the same metric {\it ansatz} utilized in this paper, and allow the vector field to have a non-zero radial and temporal component.  One can then show that the divergence of the vector field is non-zero, i.e. $\nablat_{\alpha}A^{\alpha}\neq0$, and consequently that the $K_{2}$ vector field coupling now has a non-zero contribution to the modified Einstein equations.  Therefore, the simple mapping given in (\ref{map}) can no longer hold between the original TeVeS theory and the generalized version.  Moreover, one can immediately see that when time dependence or rotation is included into the spacetime, the $K_{2}$ term will generally have non-trivial contributions due to the non-zero divergence of the vector field, and subsequently solutions of the original TeVeS theory will differ markedly from those of the generalized TeVeS theory.

Implications for observations of neutron stars and black holes given the above information is abundantly clear.  Indeed, this article implies that spherically symmetric neutron stars and black holes in the generalized TeVeS theory, where the vector field is temporally aligned, are indistinguishable from those of the original TeVeS theory.  Meanwhile, the above discussion suggests that observations of rotating black holes and neutron stars will differ between the theories, although when slow rotation approximations are valid, one would expect the observations to not differ significantly.

The relation provided in (\ref{map}) is not unique to this article.  In fact, \citet{skordis08} showed that the combination $K+K_{+}-K_{4}$ plays the same role in cosmological perturbations of the generalized TeVeS theory as ${\cal K}$ in the original theory.  Moreover, he showed that to keep the energy density of cosmological perturbations positive, the constant must satisfy $0<K+K_{+}-K_{4}<2$, which is consistent with phenomenological constraints placed of $0<{\cal K}<2$ in the original TeVeS theory \cite{sagi08,lasky08}.  This is further consistent with Einstein-\Aether theories, where this combination of constants (commonly denoted $c_{14}\equiv c_{1}+c_{4}$) must also satisfy $c_{14}<2$ such that Newton's constant is positive.

An interesting question regarding the stability of the generalized TeVeS theory now arises.  Indeed, the first blow was dealt to the stability of the original TeVeS theory when \citet{seifert07} found that the Schwarzschild-TeVeS solution \cite{giannios05} is unstable.  As the structure of the spacetime in the generalized theory is the same, does this necessarily imply that the black holes are also unstable?  It is possible that this is not the case for the following reason;
%This is probably not the case for the following reason;  
instabilities are a consequence of dynamical perturbations of the spacetime.  Therefore, while the overall structure of the background spacetime is the same for the original and generalized TeVeS theories, the field equations differ due to the extra contributions of the vector field action, and therefore the dynamics of the perturbations will also differ.  Whilst full calculations of this are still required, one can draw on experience from Einstein-\Aether theory.  There it has been shown that the vector field develops caustic singularities as it falls into the potential well of a gravitational field when the vector action is Maxwellian \cite{jacobson01}.  For this reason, generalizations of Maxwelllian vector fields are generally considered in modern Einstein-\Aether literature.  Indeed this was the original motivation for generalizing TeVeS in such a way, although I emphasize the need for these arguments to be verified in the generalized TeVeS theory by way of rigorous calculations.  

Whilst it is true that general perturbations will behave differently due to the different field equations of the theory, a certain subset of perturbations will behave in the same manner.  For example, \citet{sotani09} studied perturbations of neutron star solutions in the original TeVeS theory by assuming a Cowling approximation.  Here, the perturbation equations are simplified considerably by only dealing with perturbations of the fluid variables, and not the background spacetime (i.e. the metric, vector field and scalar field remain fixed).  As the fluid equations in the physical frame are identical for the original and generalized TeVeS theories, the results of \citet{sotani09} also hold true in generalized TeVeS, where again the mapping between the parameters is given by (\ref{map}).

In summary, in this article the equations governing spherically symmetric, static spacetimes representing neutron stars, neutral and charged black holes in the generalized TeVeS theory have been presented.  These turn out to be equivalent to those of Bekenstein's original TeVeS theory \cite{bekenstein04} given by \citet{lasky08}, \citet{giannios05} and \citet{sagi08}, where the coupling constants of the two theories are related by (\ref{map}).  Therefore, the solutions of those equations and hence the structure of the spacetimes presented by those authors is the same as for the generalized theory, and indeed these objects are observationally indistinguishable in this regime.  I have used this to show that the combination of vector coupling constants can be constrained to $K+K_{+}-K_{4}\lesssim1$ based on conservative neutron star observations.    Moreover, whilst problems with the original TeVeS theory were initiated by studies of spherically symmetric black holes, it is not expected that this will be the case for the generalized theory, given that perturbations and dynamics of such spacetimes will behave differently.

%\acknowledgments{
\vspace{0.5cm}
I thank Kostas Kokkotas, Dimitrios Giannios, Bronwen Thomas and Trixi Willburger for valuable discussions, and also the referee for valuable comments regarding the manuscript.  This work was supported by the Alexander von Humboldt Foundation by way of a Postdoctoral Research Fellowship and the Transregio 7 ``Gravitational Wave Astronomy,'' financed by the Deutsche Forschungsgemeinschaft DFG (German Research Foundation).
%}

\bibliography{GeneralizedTeVeS.Rev1}
\end{document}